# Lightweight IoT Malware Detection Solution Using CNN Classification

*Ahmad M.N. Zaza Email: az1405898@qu.edu.qa, Suleiman K. Kharroub Email: sk1404832@qu.edu.qa, Khalid Abualsaud Email: k.abualsaud@qu.edu.qa*
*Qatar University-Department of Computer Engineering and Computer Science*

*Abstract*— Internet of Things (IoT) is becoming more frequently used in more applications as the number of connected devices is in a rapid increase. More connected devices result in bigger challenges in terms of scalability, maintainability and most importantly security especially when it comes to 5G networks. The security aspect of IoT devices is an infant field, which is why it is our focus in this paper. Multiple IoT device manufacturers do not consider securing the devices they produce for different reasons like cost reduction or to avoid using energy-harvesting components. Such potentially malicious devices might be exploited by the adversary to do multiple harmful attacks. Therefore, we developed a system that can recognize malicious behavior of a specific IoT node on the network. Through convolutional neural network and monitoring, we were able to provide malware detection for IoT using a central node that can be installed within the network. The achievement shows how such models can be generalized and applied easily to any network while clearing out any stigma regarding deep learning techniques.

## I. INTRODUCTION

Internet of Things (IoT), is a term used to refer to the collection of widely internet connected devices that share data autonomously and on request over the network they are connected to. There are multiple different types of IoT devices, some of which can sense and collect data from the physical world such as temperature, and others can act upon such data in order to carry out processes that can even alter the characteristics of the physical world if need be i.e. changing the temperature of a room. IoT devices are becoming more involved in critical processes which arises the need for involving high security measures that need constant attention to function properly and there is no doubt that availability of 5G will cause more involvement in more day-to-day fields.

IoT is involved in many applications that supports many fields such as agriculture, smart power grids, smart cities, smart houses and smart vehicles. Usually there are various types of IoT devices that are involved in such applications, each with a different manufacturer. Developers tend to buy IoT devices from those manufacturers in large quantities overlooking the fact that some of those devices were designed without having security in mind which can create an array of problems in developed applications and their respective networks that they run on. The main reason behind ignoring security when manufacturing these devices is because there is a lack in security standards for IoT devices or rather enforcement of it.

In [1] and [2], it is mentioned that there are standards with regards to IoT security in both the public and the industry. Those standards are used to achieve interoperability, for regulatory compliance, compliance with public trending, and for certification purposes. However, it is also mentioned in the papers that such standards are in fact standards defined by a diverse range of industry associations across the IoT ecosystem. There is a difficulty in creating a baseline standard for IoT security that covers all IoT applications and domains. And there is another difficulty in adopting said standards, implementing them, and monitoring their effectiveness over the different IoT domains.

There are efforts in the research field to create a standard for IoT security that can cover all IoT domains and applications. ISO/IEC 27030 and ISO/IEC 30141 are an example of those efforts, which are used to assess an IoT framework and enhance its security performance. ISO/IEC 27030 and ISO/IEC 30141 dives into the trustworthiness, or in other words, how safe, secure, private, resilient, and reliable the IoT system is in the context of the reference IoT Architecture in [1].

Governments and regulatory agencies in both the US and EU are looking forward to considering the promotion of a robust IoT security standard. There are two main issues facing researchers with the task of developing a general security standard for IoT. First challenge being the diverse scope of IoT applications and their suggested standards, making it hard to find common ground between all suggested standards. Second challenge is the lack of information about implementation, review rates and the adoption of suggested standards making it hard to monitor their impact.

There is also an increase in research interest towards obtaining low latency and increased device connectivity according to [3], [4], and [5]. To achieve the trifecta of eMBB, mMTC, and uRLLC in some sense requires applying such standards mentioned previously in a correct manner to achieve reliability and low latency. Currently, 5G lacks standards that establishes what is considered as good practice or not. Which raises the question in terms of how research can contribute in implementing a concrete and verifiable way to implement IoT security.

Although there are lots of efforts in the IoT security field, it remains uncertain what is the best technology to use in order to provide maximum security without compromising on energy consumption. Since IoT devices are usually very limited in energy and computing power, so it is a challenge to come up with a low weight, high security solution.

There are multiple challenges that are present in the IoT security field ranging from data integrity, encryption capabilities, general security, patching, automation, common framework and privacy issues. The research field in this area is full of opportunities and a single project or research cannot simply address all these issues at once.

Therefore, in this research, we narrowed down the scope of our work to network level packet filtering and malware detection based on Convolutional Neural Networks (CNN).

Such approach will allow us to identify malicious nodes that are connected on a network and then enforce mitigations on such nodes so that they can co-exist safely in the same environment that has secure IoT devices. Our motivation can be seen clearly after the literature review that shows what needs to be done to provide an optimal malware detection solution for IoTs.

The reminder of the paper will be divided into sections as will be mentioned. Section II will go through background information and related works in the literature. Section III will be discussing the methodology behind our approach in details. Section IV will go over the setup we used to carry out our methodology. Section V contains the results of our experiment with comments about the outcome. Section VI is a conclusion about the experience in its entirety and what can be improved upon in the future.

## II. RELATED WORKS & BACKGROUND

In this section we will highlight recent and fundamental approaches that utilize different technologies in order to satisfy different aspects of IoT security requirements.

As we mentioned in the introduction, IoT devices require sophisticated techniques in order to provide an efficient and acceptable level of security in multiple aspects of IoT security. If we want to achieve a safe environment for IoT devices, we need to look at how we can establish such type of environment without compromising on the reliability of IoT networks.

Let's have a quick overview of what kind of QoS metrics do IoTs need to follow to achieve functional capabilities in the first place. As mentioned in [6], Quality-of-Service (QoS) is a metric that is present in almost all systems and applications. QoS refers to the efficiency and reliability of a system and it is very important to have high standards with regards to QoS. Parameters of high QoS standards are different in each system. In our case, IoT networks generally should include the following qualities: Throughput, Delay, Packet Loss, and Security.

However, it is not an easy task to meet high QoS standards in IoT based networks and systems specially when we talk about 5G networks. That is due to core features of IoT networks that enforce limitations like Resource Constrains, Platform Heterogeneity, and Dynamic Network Topology. Covering all the QoS aspects is a major need to provide a reliable implementation of security that can cover all the points above.

Next, we look at some of the solutions that we investigated going into this research. In [7], IoT security aspects were achieved using blockchain based Secure IoT control scheme. In this solution they relied on the characteristics of the blockchain, which has built-in resistant to data modification.

Although blockchains indeed provide a strong level of integrity to IoT networks, they tend to be very process heavy. In the case of IoT devices, they are very limited in terms of energy and processing power.

Paper [8] provides a different type of solution for Android IoT devices security specifically. Their main motivation came from the fact that recently, hackers are extensively exploiting the Android platform that is used in Android IoT devices which creates the challenge of securing such devices from the constantly evolving malware activities. They utilize a fusion of both blockhains and machine learning techniques to improve malware detection for Android based IoT devices.

Firstly, the machine learning automatically extracts malware information using clustering and classification technique, then save the information on the blockchain. In that case, every malware information is stored in the blockchain ledger and is visible to all nodes on the network. Ledger capabilities are then used to store malware related data which can then be extracted at later time.

The downfall of the two previous papers is the lack of awareness to the effects of the implementation on QoS of the network. Having heavy workload to ensure security is counterproductive when it comes to using such techniques with IoT devices, thus making such implementations almost obsolete.

In [9], a lightweight classification model was developed specifically for IoT Malware detection based on Image recognition which helps in what we are doing. Having a lightweight classification model is very important, especially if we are working with IoT devices with limited energy and computational power. In terms of methodology, they focus on mitigating only DDoS malware in IoT environments. By extracting images of malware which are grey scale images converted from a malware binary. A lightweight CNN is utilized to classify the family of each malware based on the malware image. the solution provides 94% accuracy for goodware classification and DDoS malware, and 81.8% accuracy for goodware classification and two main malware families. They use a dataset acquired from the work found in [10] which focuses mainly on collecting IoT malware that was used to test how well the system performs.

- ***Research Motivation:***

Our motivation behind choosing CNN is that it provides a lightweight solution for malware detection which can be easily applied to IoT networks that might be using 5G connectivity. There is a common misconception regarding CNN or machine learning techniques in general, which is thinking that such techniques are very process heavy and require lots of resources in order to work. This misconception is half true since training the model is very time consuming and process heavy. But once we have our trained model, a lightweight model is produced. With the aid of a centralized node, any files that are communicated on our IoT network can go through if needed, which will classify these files as goodware, or malware while also specifying its category as we will see later. The fact that once the malware detection model is trained, we can easily deploy it over IoT networks since it becomes very light weight after training is done.

## III. METHODOLOGY

The approach we are taking in this experiment is that we need to establish certain guidelines in order to achieve a reasonable outcome out of the model we are developing. This includes multiple steps that we will be discussing here but let's look first at the main contributions of this paper:

- Looking at a bigger dataset than any of the other papers available
- Discuss new ways of implementing CNNs
- Show outperformance to other solutions in terms of time and efficiency and resources while only providing binary categorization while ours has multi-categorial classification
- Test against Obfuscation Techniques for IoT apps
- Provide a concrete implementation that can be deployed in real life

With that in mind, the process we hope will be able to be generalized and implemented easily for those who are looking to improve upon or fully integrate the solution here.

Our methodology includes looking at certain steps that will tremendously improve the outcome of a classifier especially when using CNN as the main model for classification. The proposed solution will have to include a lightweight CNN model of which will have to handle different scenarios using different data representation techniques of which we will test and measure performance metrics to provide results and conclusions regarding what should be done. The model will run on the "application" layer which will allow us to access binary files/packets going through the network for classification as well as "network" layer if we have access o it hypohtically.

The dataset we have acquired included the malware that was collected by Professor Katsunari Yoshioka from IoTPoT [10] of which he has wrote a paper about regarding the assembly of this dataset. We were able to get over 5000 samples of multiple IoT malware that we will be using for training the model. VirusTotal API [11] was used to label all the acquired dataset and thus establishing base truth that we will conduct experiments on. There were many different variants of the same viruses but after a deeper look we found out that most of the samples were part of the Linux.Gafgyt or Linux.Mirai family. This was the latest acquired dataset of IoT malware acquired from a honeypot. The dataset was inclusive of very light malware which can only propagate through IoT devices which limited the variety of malicious binary files obtained. This has made classification and categorization much simpler and this is what we will separate our malwares into for the sake of convenience and consistency. As for goodware, we collected around 1000 samples of Linux based software as most IoTs use Unix based kernels and added it to the overall collection.

Now comes the data representation part and we have decided to go with 2 main interpretations of the binary files we have acquired. First approach was to read the binary files in bytes and convert each byte into a value between 0 and 255, producing a grayscale image in the process as seen in Fig. 1.

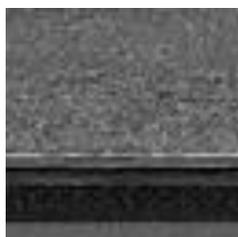

*Figure 1. grayscale image representation of Virus of md5 hash of 1ea2c4bd2717598a96bba94ac5da8f71*

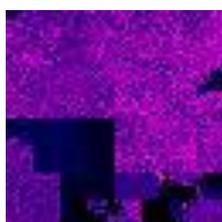

*Figure 2. Hilbert curve (Entropy) image representation of Virus of md5 hash of 1ea2c4bd2717598a96bba94ac5da8f71*

Second way of representing data was using the Hilbert curve of which an entropy cluster image is produced. We found out according to [12] that the Hilbert curve is a remarkable construct in many ways, but the thing that makes it valuable in the field of computer science is that it has good clustering properties. This combined with CNN can give us better results as we need to cluster features in a better manner for the model to easily be able to differentiate inputs and make correct predictions. The representation of data also considers reading the bytes of the file and then passing them through a function that generates the picture presented in Fig.2 acquired through [13].

We will use different input matrices in this case of which will affect the CNN model in terms of how the code will change but not the overall outcome or procedure that will be used in the experiments conducted.

Now comes the part of building the CNN model. To build a reliable model, we need to optimize many of the parameters and see how that will affect the outcomes and the performance of the model in order to make a generalized model that can fit the scenarios we face while classifying malware for IoT devices.

Previous work in [9] has only used two convolutional layers without looking further into optimizing the outcome of the model. We will further expand on that and test all metrics to understand what needs to be done within this problem we have that is multi-categorial classification.

```
Model: "sequential"

Layer (type)                 Output Shape              Param #
=================================================================
conv2d (Conv2D)              (None, 64, 64, 64)        128
max_pooling2d (MaxPooling2D) (None, 32, 32, 64)        0
conv2d_1 (Conv2D)            (None, 32, 32, 128)       8320
max_pooling2d_1 (MaxPooling2 (None, 16, 16, 128)       0
conv2d_2 (Conv2D)            (None, 16, 16, 128)       16512
flatten (Flatten)            (None, 32768)             0
dense (Dense)                (None, 128)               4194432
dense_1 (Dense)              (None, 3)                 387
=================================================================
Total params: 4,219,779
Trainable params: 4,219,779
Non-trainable params: 0
```

*Figure 3. CNN Model Summary*

Our CNN model shown in Fig. 3 added an extra convolutional layer of which will help in further improving the accuracy. You can also see that we are expanding the feature rather than reducing them as we already have little sized inputs that we are generating to accommodate the limited resources that are in IoTs. ReLU activation function was used as it is the most reliable gate function available for such cases.

We have decided on the size of the input and the many other parameters including: image_size, batch_size, epochs, loss function, and other metrics to consider based on different experiments we conducted to find out what would work best in a computationally hindered networks and devices. We will now go over them in the next section which will explain all the numbers and figures acquired in all details.

IV. EXPERIMENT SET-UP

We have gone over several parameters as we discussed, and we will explain all the outcomes we have found here. First, we need to establish the environment we have ran our experiments on as well as the specifications to give a relative measure of how our work was computed against. Hardware specifications are as follows:

**CPU Architecture and OS:** Windows 10 Pro with Intel i7-7700HQ – 8GB RAM **GPU Architecture:** NVIDIA GeForce GTX 1050 Max-Q (2GB RAM)

As for software specifications, these are the following programs we used:
- Python => 3.7+
- TensorFlow => 2.0+
- matplotlib.pyplot
- PIL (Python Image Processing Library)

It can be noticed that our implementation is quite simple and easy to reproduce using very few programs and lines of code which makes our approach highly scalable if needed.

The approach we followed consists of three steps. These steps include data initialization and representation, training of the model using a training set, and testing several scenarios and optimize based on results. We will repeat these 3 steps until we achieve the desired results we need.

We have a look at data initialization and how the different data representations affected that time. Table I shows the different timings that the two different methods had on initialization time on average in any given run. We can see a huge difference in timings of which is a critical point to consider when it comes to implementing such classifier in an IoT environment. We will discuss whether this tradeoff is worth it or not in later parts in this section. For now, we will run all our testing on grayscale data image vectors for training the model as both have same image output size in all test runs.

*TABLE I. Initialization Timings for Different Data Representations*

| Technique (64x64 image production) | Avg. Time (Seconds) |
|---|---|
| Byte Reading | 6.7 |
| Entropy Generation | 2514 |

As we established previously, categorization of software included the Gafgyt and Mirai families and those that did not belong to either were the samples of goodware labeled as "gooodware". So, we ran a simple CNN model using these classes as predictions with different image sizes that we have produced to see the effect on processing time. Another observation is that the time increases exponentially as the number of inputs are squared because we are inserting a 2D array of values which is reflected upon time as well. 64x64 images were processed almost in square root time of images of size 128x128 (15s vs. 220s respectively)

We have chosen to go with 64x64 images as it gives us the most reliable time with no compromise in decreasing accuracy when training the model. The balance of performance is what we are trying to do when optimizing all the parameters.

Now comes epochs, or iterations of training, to see how the model we have for the given dataset behaves. According to [15], in the case of one epoch, an entire dataset is transmitted forward and backward through the neural network once only. We divide the epoch into several smaller batches to be trained, since one epoch is relatively too big.

The default batch size is 32 in TensorFlow and so we ran different number of epochs on the dataset we have initialized. We found that the epochs curve flattens for accuracy after 4 or 5 iterations and that would be our maximum number of iterations to save time and energy. Batch size will not affect the time nor the accuracy as it is relative to the dataset and how randomized the data is, so we went with batches of 32 for training.

The loss value we were trying to optimize was 'sparse_categorical_crossentropy' with 'adam' as the optimizer function [16][17]. Machines learn through the use of a loss function. It's a method of evaluating how well specific algorithm models the given data. If the predictions deviate too greatly from the actual results, loss function would produce a very large number. The optimizer and loss value best describe the problem we are trying to tackle in order to differentiate multicategories of malware along with goodware. With the help of some optimization function, loss function gradually learns to reduce the error in prediction. We used default parameters for the functions we have used and have achieved a loss value of below 0.1 on average. And from the data acquired, validation accuracy has met the accuracy of the model indicating that that there was no overfitting in the data we have trained and acquired.

As for the architecture, Fig. 4 shows a suggested way of developing an implementation which can help in providing proper classification for a network with IoT devices. Models can be trained either on the edge or in the cloud, if it needs further processing power, to establish a more concrete model that can give us high accuracy all the time.

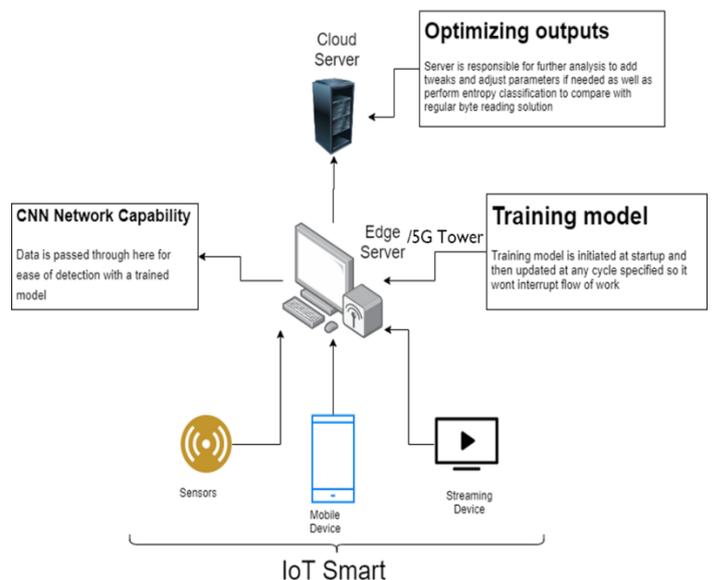

*Figure 4. Our Suggested Architecture the System Will Work In*

| Class | TP | FP | FN |
|---|---|---|---|
| Mirai | 0.923 | 0.035 | 0.042 |
| Gafgyt | 0.958 | 0.033 | 0.009 |
| Goodware | 0.987 | 0.01 | 0.003 |
| Average | 0.956 | 0.026 | 0.018 |

| Class | Recall | Precision |
|---|---|---|
| Mirai | 0.956477 | 0.963466 |
| Gafgyt | 0.990693 | 0.9667 |
| Goodware | 0.99697 | 0.98997 |
|  | 0.98138 | 0.973379 |
| F-1 Score overall | 0.977363 | |

*Figure 5. Comparison Tables of Relatable Scores and Metrics*

## V. RESULTS

After we set up the data and the model, the experiment parameters were as follows:

- 5000 malware samples
- 1000 goodware samples
- Variable size of training sets
- Epochs = 5
- Three Convolutional layers
- Three pooling layers
- Three categories of classification
- Grayscale Images (64, 64, 1)
- Time per batch of 32 ~= 0.05 seconds
- No. of total runs: 1000 times
- All numbers represent averages
- Inputs are all randomized

The malicious nodes can be any of the participating devices in the network. We assume we already gathered the binary files or packets that contain malicious code or scripts. We ran with grayscale images here for time appropriation as the running time for both data representations are both similar. We have run those variables on different categories and seen the effects of changing certain aspects and how they affect the accuracy and timings.

After we have had the initial runs and seen what the amount of training data is needed to be in the set, we have chosen 70% of the samples we had for training randomly and recorded Precision, Recall, and F-1 Scores averages. In Fig.5, we can see the comparison of different scores in tables compared to other relevant solutions especially found in [9]. Our solution provides a much higher accuracy and much less malware identified as benign. It is very dangerous for having a false positive of goodware because it can put the whole network at risk especially in sensitive applications such as real time networks.

With F1-Scores being shown, the relatively small change we have introduced to the models presented by others has gained a lot of significant improvement as the metrics have shown. This is still lightweight and very compact with very minimal computational resources that have given such amazing performance.

As for the entropy representation of data, Fig. 6 shows a comparison between them. The input changed to an RGB value of (64x64x3) and the CNN model changed accordingly. Timing wise we did not see a huge difference in training duration and thus did not include it here.

As for accuracy, the entropy representation gave high accuracy with a rather smaller dataset than grayscale images. This can help in achieving better accuracy with smaller data collected if need and that shows the power of it to maybe improve the accuracy if ran on a cloud node to see how well we can optimize the data by generating even smaller models through this technique. We will discuss further discussions at the end of this section.

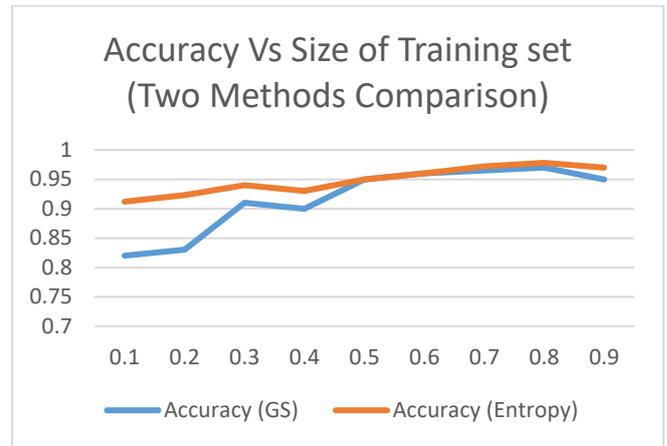

*Figure 6. Accuracy Vs Size of Training set (Two Methods Comparison)*

After we have evaluated the model, we tried obfuscated code and have tried to see how this affects the overall integrity of our model. A lot of attackers try to obfuscate their code if they want to surpass certain classification analysis to overthrow the model into thinking that they don't belong to a certain malware family. Despite having many papers such as [9], [14] say that attackers on IoT devices tend not to obfuscate their code, we have anyway tried to see what the outcome would be otherwise.

Obfuscators usually apply transformations directly on the application's bytecode. Since there is no readable equivalent, some of these bytecode changes cannot be reverted back to source code. We test several Android APKs with obfuscated code (generated by GuardSquare Software) [18] on our grayscale data initialization method. We are not using any disassembler but rather just reading bytes directly contrary to decompiling which can overcome the obfuscation methods point of impact.

We have run a different sample of a 1000 APKs to see how our model would perform and have chosen the android platform as most smart devices nowadays run on it. The two different methods of data representation and the results are displayed in Fig. 7. The accuracy has dropped initially for smaller datasets compared to the IoT binaries samples as it seems the obfuscation does alter the behavior of data reading in our implementation. However, we can see we can still achieve ~92% accuracy on average despite that. With proper modifications to the model, this can easily intercept obfuscated malware with proper optimization while also interfering attacks for which the signature is not known.

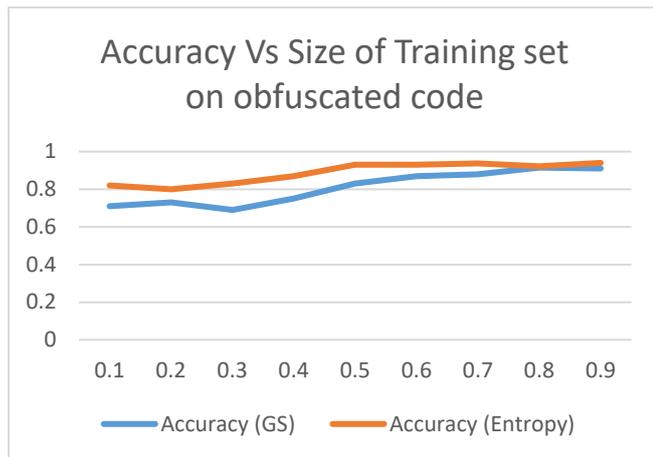

Figure 1. Accuracy Vs Size of Training set on obfuscated code

The focus of this study involves the optimization of a CNN model for the sake of producing an optimal design that can help in a unique way of identifying malware. With a relatively small HDF5 file being produced, the model can easily be propagated after being trained remotely despite already having small durations of initializing and training against samples of data. Limits on such files produced according to [19] are quite minimalistic and can be used on multiple IoT devices and platforms through simple UIs if needed for the user to interact with. This is a holistic approach to the whole problem of data classification in resource limited devices and hardware with very inexpensive gear to run on.

As for the power usage using original inputs, the GeForce GTX 1050 Max-Q is rated at 34 to 40 Watt [20]. That is very low wattage while giving decent learning times as well. the average percentage of utilization of the machine used during the learning phase was almost only 30% of the GPU power is being utilized while the CPU was on idle as it was managing some other background processes. This shows that this implementation is not as power hungry as other methods used due to the simplistic model we are producing.

We weren't able to use tools to measure exact power drain for the GPU usage as this is a mere simulation, but future work will look into implementing this with a working remote IoT network while having a full-scale implementation running to showcase the true capability of the system proposed. Energy efficient protocols can be combined with the lightweight model generated to provide updates periodically to the system. We suggest looking at 6LoWPAN or ZigBee protocols to ensure low power usage which hopefully should be included in upcoming studies we might conduct. Packet filtering can also be another application which can benefit in the classification of malicious data sent over the network. The application covers a wide range of areas that were not considered before specifically for IoT.

Future work can include further optimization of the dataset itself as well as understanding the unique attacking techniques the malicious users must establish more parameters that can further enhance the performance of the model. Another noteworthy point to explore is the application of Generative Adversarial Networks (GANs) and how they can help in predicting future malware and seeing if that would be an outstanding move to further demean attacks from happening. More real-life application and data gathering of the architecture is in progress to see latency and overhead effect on the performance of the model and see how simulation vs practical can differ from the results found here. Also, IoT devices will be deployed in a large field for long times, the update requirement is critical so how often should the CNN be retrained and how is it going to be established and what is the risk of not updating the classifier should be a critical point for consideration.

## VI. CONCLUSION

As the IoT devices numbers are on the rise, so is the exploitation of them for malicious reasons and thus the raise in need for an efficient way of detecting such malicious acts. A lightweight and resource friendly solution is needed to cater for the very limited devices that we are carrying around all the time which are susceptible to many dangers. CNN comes to solve this problem by introducing deep learning techniques to identify hidden features and patterns that can aid in detecting malware. We have provided a concrete and generalized way to understanding the problem through the CNN implementation and having easy and efficient programs that can withstand and uphold to the task at hand of quick and precise classification. The architecture and model we have provided has proven to be efficient in both time and accuracy while also being considerate of the limited resources that IoTs have. After studying the parameters and providing the building blocks for a deployment ready model, we had and outstandingly high overall accuracy and many ways to detect malware even if the code is obfuscated. We have proven that simple techniques can have huge effects on the classification issue presented in IoT networks of which can provide a safe environment for all kinds of scenarios and applications. The solution that we provided in this paper can be useful in multiple IoT applications like agriculture and drone surveillance system that depend on 5G connectivity.

ACKNOWLEDGMENT

This work was made possible by NPRP grant # 10-1205-160012 from the Qatar National Research Fund (a member of Qatar Foundation). The statements made herein are solely the responsibility of the authors.

REFERENCES

[1] Franberg, O., (2019). "*Preparing The ISO/IEC 30141 Iot Reference Architecture Edition 2 Through A Mindshare*," 20-Jun-2019. [Online] Available at: <https://iotweek.blob.core.windows.net/slides2019/4.%20THURSDAY%2020/IoT%20Systems%20Architectures%2C%20Models%2C%20Guidelines/20190620%20IoT%20Week%20session%20architecture%20SC41%20OstenFramberg%20-%20AK%20v2.pdf> [Accessed 20 April 2020].

[2] M. A. Al-Garadi, A. Mohamed, A. Al-Ali, X. Du, I. Ali, and M. Guizani, "A Survey of Machine and Deep Learning Methods for Internet of Things (IoT) Security," *IEEE Communications Surveys & Tutorials*, pp. 1–1, 2020.

[3] S. Li, L. D. Xu, and S. Zhao, "5G Internet of Things: A survey," *Journal of Industrial Information Integration*, 20-Feb-2018. [Online]. Available: https://www.sciencedirect.com/science/article/abs/pii/S2452414X18300037. [Accessed: 29-May-2020].

[4] M. Agiwal, A. Roy and N. Saxena, "Next Generation 5G Wireless Networks: A Comprehensive Survey," in *IEEE Communications*

*Surveys & Tutorials*, vol. 18, no. 3, pp. 1617-1655, thirdquarter 2016, doi: 10.1109/COMST.2016.2532458.

[5] L. Chettri and R. Bera, "A Comprehensive Survey on Internet of Things (IoT) Toward 5G Wireless Systems," *IEEE Internet of Things Journal*, vol. 7, no. 1, pp. 16–32, 2020.

[6] Zaza, A., Al-Emadi, S. and Kharroub, S., (2020). "Modern QoS Solutions in WSAN: An Overview of Energy Aware Routing Protocols and Applications." In: 2020 IEEE International Conference on Informatics, IoT, and Enabling Technologies (ICIoT'20). Doha, Qatar.

[7] Choi, S., Burm, J., Sung, W., Jang, J. and Heo, Y. (2018). "*A Blockchain-based Secure IoT Control Scheme*." In: 2018 International Conference on Advances in Computing and Communication Engineering. France.

[8] Kumar, R., Zhang, X., Wang, W., Khan, R., Kumar, J. and Sharif, A. (2019). "*A Multimodal Malware Detection Technique for Android IoT Devices Using Various Features*." IEEE Access, 7, pp.64411-64430.I.

[9] Su, J., Vargas, D., Prasad, S., Sgandurra, D., Feng, Y. and Sakurai, K. (2018). "*Lightweight Classification of IoT Malware Based on Image Recognition*." In: 42nd IEEE International Conference on Computer Software & Applications.

[10] Pa, Y., Suzuki, S., Yoshioka, K., Matsumoto, T., Kasama, T. and Rossow, C., (2020). "IoTPOT: analysing the rise of IoT compromises." EMU, 9, p.1.

[11] "virustotal-api," *PyPI*. [Online]. Available: https://pypi.org/project/virustotal-api/. [Accessed: 20-Apr-2020].

[12] "binvis.io - a browser-based tool for visualising binary data," *cortesi - binvis.io - a browser-based tool for visualising binary data*. [Online]. Available: https://corte.si/posts/binvis/announce/index.html. [Accessed: 26-Nov-2019].

[13] Cortesi, "cortesi/scurve," *GitHub*. [Online]. Available: https://github.com/cortesi/scurve/blob/master/binvis. [Accessed: 20-Apr-2020].

[14] Nguyen, Khanh Duy Tung, et al. "Comparison of Three Deep Learning-Based Approaches for IoT Malware Detection." *2018 10th International Conference on Knowledge and Systems Engineering (KSE)*, 2018, doi:10.1109/kse.2018.8573374.

[15] Sharma, Sagar. "Epoch vs Batch Size vs Iterations." *Medium*, Towards Data Science, 5 Mar. 2019, [Online] towardsdatascience.com/epoch-vs-iterations-vs-batch-size-4dfb9c7ce9c9. [Accessed: 20-Apr-2020].

[16] *Optimizers - Keras Documentation*. [Online]. Available: https://keras.io/optimizers/. [Accessed: 20-Apr-2020].

[17] "Module: tf.losses : TensorFlow Core r2.0," *TensorFlow*. [Online]. Available: https://www.tensorflow.org/api_docs/python/tf/losses. [Accessed: 20-Apr-2020].

[18] "Code hardening (obfuscation & encryption)," *Guardsquare*, 20-May-2019. [Online]. Available: https://www.guardsquare.com/en/mobile-application-protection/code-hardening-obfuscation-encryption. [Accessed: 20-Apr-2020].

[19] *Limits in HDF5*. [Online]. Available: https://support.hdfgroup.org/HDF5/faq/limits.html. [Accessed: 26-Nov-2019].

[20] K. Hinum, "NVIDIA GeForce GTX 1050 Max-Q GPU," *Notebookcheck*, 12-Jan-2018. [Online]. Available: https://www.notebookcheck.net/NVIDIA-GeForce-GTX-1050-Max-Q-GPU.277746.0.html. [Accessed: 21-Apr-2020].